%% file: main.tex
\documentclass[ALICE,manyauthors]{cernphprep}
\usepackage[comma,square,numbers,sort&compress]{natbib}

\usepackage{lineno}
\usepackage{xspace}
\usepackage{hyperref}
\usepackage[T1]{fontenc}
\usepackage{color}

\begin{document}
\input{commands.tex}

\begin{titlepage}
\PHyear{2021}       
\PHnumber{088}      
\PHdate{10 May}  

\title{Charm-quark fragmentation fractions and production cross section at midrapidity in pp collisions at the LHC}
\ShortTitle{Fragmentation fractions and charm production cross section}   

\Collaboration{ALICE Collaboration\thanks{See Appendix~\ref{app:collab} for the list of collaboration members}}
\ShortAuthor{ALICE Collaboration} 

\begin{abstract}

Recent \pt-integrated cross section measurements of the ground-state charm mesons and baryons, \Dzero{}, \Dplus{}, \Ds{}, \Lc{}, and \XicZero{}, are used to evaluate the charm fragmentation fractions and production cross section per unit of rapidity at midrapidity ($|y|<0.5$), in pp collisions at $\s = 5.02~\TeV$ at the LHC. The latter is $\de\sigma^{\ccbar}/\de y|_{|y| < 0.5}=1165\pm 44(\mathrm{stat})^{+134}_{-101}(\mathrm{syst})~\mub$.
These measurements were obtained for the first time in hadronic collisions at the LHC including the charm baryon states, recently measured by ALICE at midrapidity. 
The charm fragmentation fractions differ significantly from the values measured in \ee{} and ep collisions, providing evidence of the dependence of the parton-to-hadron fragmentation fractions on the collision system, indicating that the assumption of their universality is not supported by the measured cross sections. An increase of a factor of about 3.3 for the fragmentation fraction for the \Lc{} with a significance of $5\,\sigma$ between the values obtained in pp collisions and those obtained in \ee (ep) collisions is reported. The fragmentation fraction for the \XicZero{} was obtained for the first time in any collision system.
The measured fragmentation fractions were used to update the \ccbar{} cross sections per unit of rapidity at $|y|<0.5$ at $\s{} = 2.76$ and 7 TeV, which are about 40\% higher than the previously published results.
The data were compared with perturbative-QCD calculations and lie at the upper edge of the theoretical bands.

\end{abstract}
\end{titlepage}

\setcounter{page}{2} 

The study of heavy-flavour hadron production in proton--proton (pp) collisions provides an important test for quantum chromodynamics (QCD) calculations. The transverse-momentum (\pt) differential cross sections of charm mesons measured in pp collisions by the ALICE~\cite{Abelev:2012vra,Abelev:2012tca,Acharya:2017jgo,Acharya:2019mgn,Acharya:2021cqv}, ATLAS~\cite{Aad:2015zix}, CMS~\cite{Sirunyan:2017xss}, and LHCb~\cite{Aaij:2013mga,Aaij:2015bpa,Aaij:2016jht} experiments at the LHC and the STAR~\cite{Adamczyk:2012af} experiment at RHIC, as well as in \ppbar{} collisions by the CDF~\cite{Acosta:2003ax} experiment at the Tevatron, are described within uncertainties by perturbative-QCD (pQCD) calculations having next-to-leading order (NLO) accuracy with all-order resummation of next-to-leading logarithms, such as FONLL~\cite{Cacciari:1998it,Cacciari:2001td,Cacciari:2012ny} and NLL~\cite{Kniehl:2004fy,Kniehl:2012ti,Benzke:2017yjn,Kramer:2017gct,Helenius:2018uul}. These calculations are based on the factorisation theorem, according to which the \pt-differential cross sections are computed as the convolution of three terms: (i) the parton distribution functions (PDFs) of the incoming (anti)protons, (ii) the partonic cross section, calculated as a perturbative series in powers of the strong coupling constant $\alpha_\mathrm{s}$, and (iii) the fragmentation functions which describe the transition from charm quarks into charm hadrons. The latter, in these calculations, are typically parametrised from measurements performed in \ee{} or ep collisions~\cite{Braaten:1994bz}, under the assumption that the hadronisation of charm quarks into charm hadrons is a universal process independent of the colliding systems. Accordingly, measurements of charm mesons were exploited in the past  to derive a measurement of the charm production cross section at hadron colliders, by scaling the production cross section of the D mesons with the corresponding charm-quark fragmentation fraction, $f(\mathrm{c\to D})$, taken from \ee collisions~\cite{Abelev:2012vra,Acharya:2017jgo,Aaij:2015bpa,Aaij:2016jht,Adamczyk:2012af,Gladilin:2014tba}. 

Recent measurements of charm-baryon production at midrapidity in pp collisions showed an enhancement of the $\Lc/\Dzero$~\cite{Acharya:2017kfy,Acharya:2020uqi,Acharya:2020lrg,Sirunyan:2019fnc} and $\XicPlusZero{}/\Dzero$~\cite{Acharya:2017lwf,Acharya:2021dsq,xic13tev} ratios for $\pt < 6-8~{\rm GeV}/c$ with respect to the ones measured in \ee collisions. These measurements suggest a significant difference of the fragmentation fractions of charm quarks into charm baryons in hadronic collisions
at LHC energies compared to those measured in \ee{} and ep collisions. These findings are similar to those obtained in the beauty sector by the CDF Collaboration at the Tevatron~\cite{Aaltonen:2008zd} and by the LHCb Collaboration at the LHC~\cite{Aaij:2011jp,Aaij:2019pqz}. 
Several models based on different assumptions, like the inclusion of hadronisation via coalescence~\cite{Song:2018tpv,Minissale:2020bif}, or considering a set of yet-unobserved higher-mass charm-baryon states~\cite{He:2019tik}, or including string formation beyond the leading-colour approximation~\cite{Christiansen:2015yqa}, have been proposed to explain the baryon enhancement. Updates of the fit to the measured fragmentation functions of ${\rm c} \rightarrow \Lc{}$ in \ee collisions were also performed~\cite{Maciula:2018iuh,Kniehl:2020szu} without improving the agreement between data and model calculations. 
These observations required a new approach for evaluating the charm-quark production cross section at midrapidity and the charm-quark fragmentation fractions based on the measurements of both charm mesons and baryons. 

The measurements described above not only provide constraints to pQCD calculations but are also important as references for the investigation of the charm-quark interaction with the medium created in heavy-ion collisions.
In particular, in the context of the heavy-ion programme at the LHC, the \ccbar{} production cross section per nucleon--nucleon collision is a fundamental ingredient for the determination of the amount of charmonium production by (re)generation in the quark--gluon plasma (QGP)~\cite{BraunMunzinger:2000px,He:2019tik,Zhao:2011cv,Liu:2009nb}, a mechanism that is supported by \Jpsi{} measurements in nucleus--nucleus collisions at the LHC~\cite{Adam:2016rdg,Acharya:2019lkh}.

In this Letter, the charm fragmentation fractions and the charm production cross section per unit of rapidity at midrapidity ($|y|<0.5$) in pp collisions at $\s = 5.02~\TeV$ are reported. The results were obtained by considering the contribution based on the measurement of the ground-state charm hadrons \Dzero{}, \Dplus{}, \Ds{}, \Lc{}, and \XicZero{} by the ALICE Collaboration~\cite{Acharya:2021cqv,Acharya:2021dsq,Acharya:2020uqi}. 

The ALICE experiment and its performance are presented in detail in~\cite{Aamodt:2008zz,Abelev:2014ffa}. The main detectors used for the measurements presented here are the Inner Tracking System, the Time Projection Chamber and the Time-Of-Flight detector for vertexing, tracking, and particle identification purposes. 
The data from \pp collisions at $\s = 5.02$~TeV were collected during the 2017 run with a minimum bias trigger, and they correspond to an integrated luminosity ${L_{\rm{int}}}$= (19.3 $\pm$ 0.4)~$\rm{nb^{-1}}$~\cite{ALICE-PUBLIC-2018-014}. 
D mesons were reconstructed from their decays ${\rm \Dzero \rightarrow K^{-}\pi^+}$, ${\rm \Dplus \rightarrow K^{-} \pi^+ \pi^+}$, ${\rm \Ds \rightarrow \phi\pi^+ \rightarrow K^{-} K^{+} \pi^+}$, and ${\rm \Dstar \rightarrow \Dzero\pi^+}$, and charm baryons from their decays ${\rm \Lc \rightarrow p K^{0}_S}$, ${\rm \Lc \rightarrow p K^{-}\pi^+}$, and ${\rm \XicZero \rightarrow \Xi^{-}{\rm e}^{+}\nu_{\rm e}}$. The charge conjugates are measured as well and the results are averaged. 
The cross sections of \Dzero{} and \Dplus{} mesons were measured down to $\pt=0$~\cite{Acharya:2021cqv}. The cross sections for \Dstar{} and \Ds{} mesons were measured down to \pt = 1 GeV/$c$, corresponding to about 80\% of the integrated cross section~\cite{Acharya:2019mgn}. The \Lc{} baryon cross section was measured down to $\pt=1~\GeV/c$, corresponding to about 70\% of the integrated cross sections~\cite{Acharya:2020uqi,Acharya:2020lrg}. The \XicZero{} baryon was measured down to $\pt=2~\GeV/c$, corresponding to about 40\% of the integrated cross section~\cite{Acharya:2021dsq}.
The systematic uncertainties of the meson and baryon measurements include the following sources: (i) extraction of the raw yield; (ii) prompt fraction estimation; (iii) tracking and selection efficiency; (iv) particle identification efficiency; (v) sensitivity of the efficiencies to the hadron \pt shape generated in the simulation; (vi) \pt-extrapolation for the hadrons not measured down to $\pt =$ 0. In addition, an overall normalisation
systematic uncertainty induced by the branching ratios (BR)~\cite{Zyla:2020zbs} and the
integrated luminosity~\cite{ALICE-PUBLIC-2018-014} were considered. 

\begin{figure}[!tb]
\begin{minipage}[c]{0.5\linewidth}
\centering
\includegraphics[width=1.\textwidth]{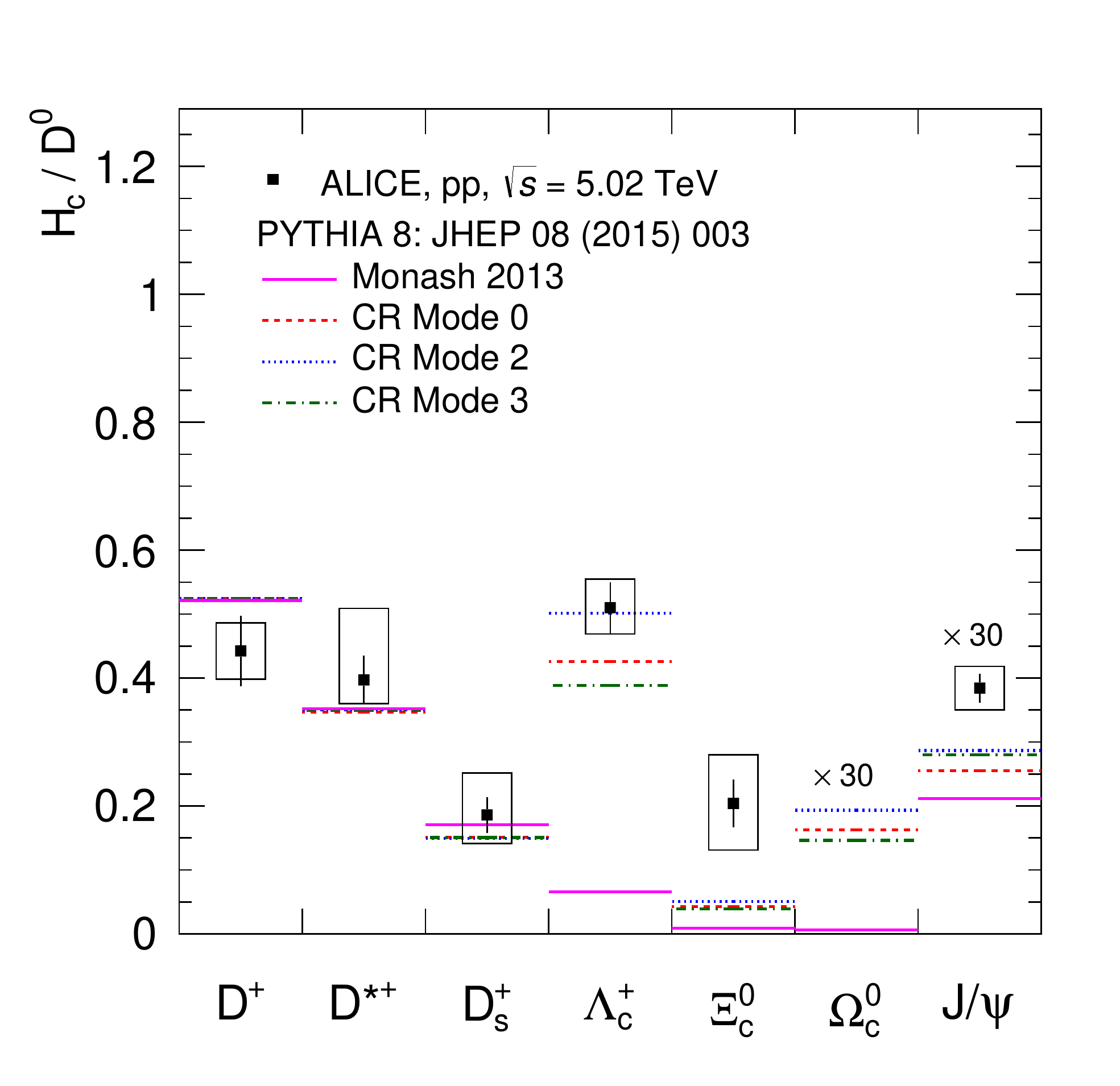}
\end{minipage}
\hspace{0.2cm}
\begin{minipage}[c]{0.5\linewidth}
\centering
\includegraphics[width=1.\textwidth]{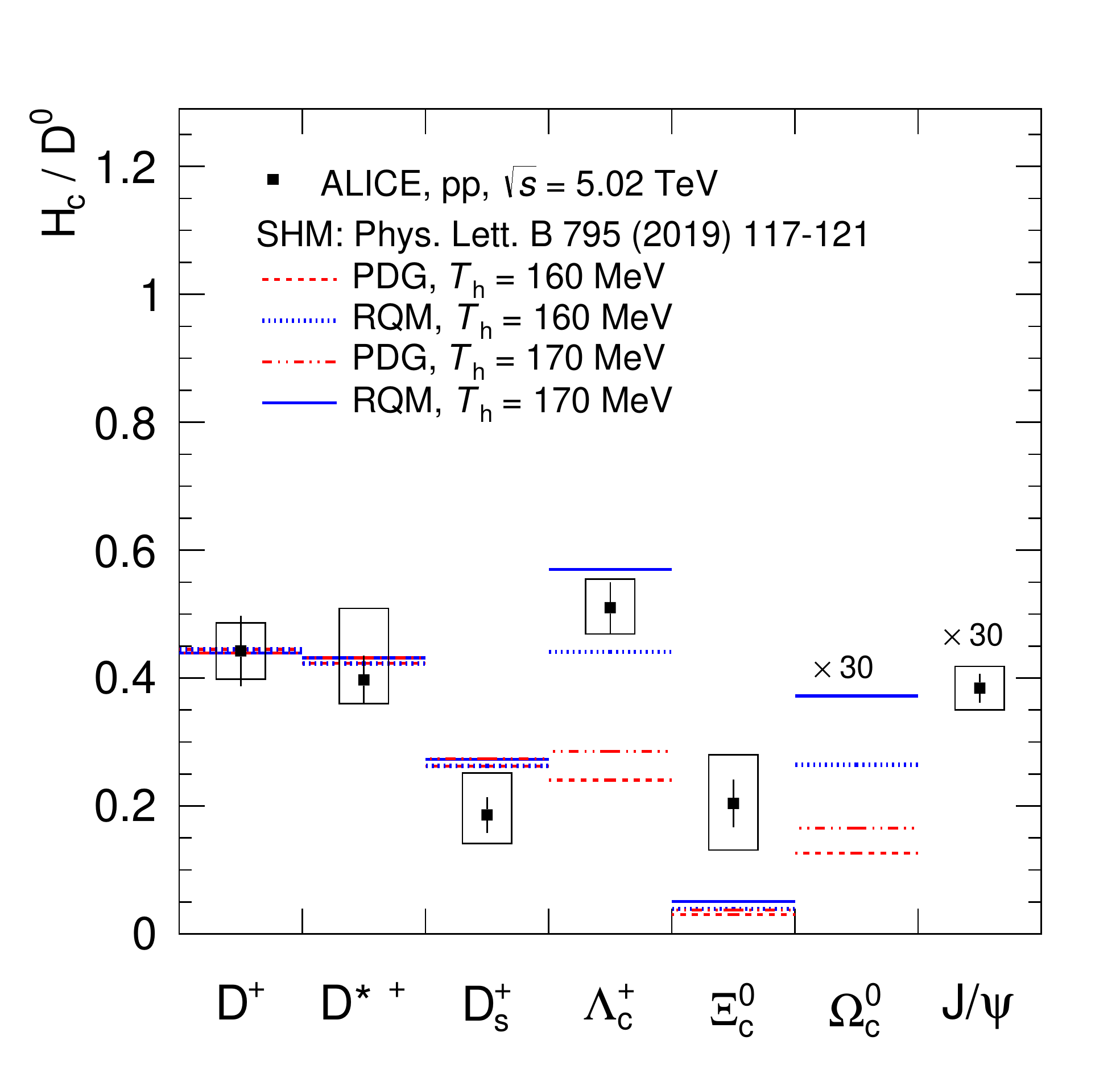}
\end{minipage}%
\caption{Transverse-momentum integrated production cross sections of the various charm meson~\cite{Acharya:2019mgn,Acharya:2021cqv,Acharya:2019lkw} and baryon~\cite{Acharya:2020uqi,Acharya:2020lrg,Acharya:2021dsq} species per unit of  rapidity at midrapidity normalised to that of the \Dzero{} meson measured in pp collisions at $\s = 5.02~\TeV$. The measurements are compared with PYTHIA~8 calculations~\cite{Skands:2014pea,Christiansen:2015yqa} (left panel) and with results from a SHM~\cite{He:2019tik} (right panel) (see text for details). For \Jpsi the inclusive cross section was used. The $\Jpsi/\Dzero$ ratio, as well as the model calculations for the $\Omegac/\Dzero$ ratio, are multiplied by a factor 30 for visibility.}
\label{figshm}
\end{figure}

Figure~\ref{figshm} shows the \pt-integrated production cross sections per unit of rapidity of the various open- and hidden-charm meson (\Dplus{}, \Ds{}, \Dstar{}, and \Jpsi{})~\cite{Acharya:2019mgn,Acharya:2021cqv,Acharya:2019lkw} and baryon (\Lc{} and \XicZero{})~\cite{Acharya:2020uqi,Acharya:2020lrg,Acharya:2021dsq} species, obtained in pp collisions at $\s = 5.02~\TeV$, as the average of particle and antiparticle, and normalised to the one of the \Dzero{} meson. When computing the ratios between the different hadron species, systematic uncertainties due to tracking, the feed-down from beauty-hadron decays, the \pt-extrapolation, and the luminosity were propagated as correlated. 
For the \XicZero{} baryons, the additional contribution to the beauty feed-down systematic uncertainty due to the assumed \Xib{}-baryon production relative to that of \Lb{} baryons~\cite{Acharya:2021dsq,xic13tev} was considered as uncorrelated with the uncertainties related to the beauty feed-down subtraction for the other charm hadron species.
In the $\Jpsi{}/\Dzero{}$ ratio all the systematic uncertainties were propagated as uncorrelated, with the exception of the luminosity uncertainty. The treatment of the systematic uncertainties is also the same for the computation of the other quantities reported here. 

In the left panel of Fig.~\ref{figshm} the experimental data are compared with results from the PYTHIA~8 generator, using the Monash 2013 tune~\cite{Skands:2014pea}, and tunes that implement colour reconnections (CR) beyond the leading-colour approximation~\cite{Christiansen:2015yqa}.
In the Monash 2013 tune, the parameters governing the heavy-quark fragmentation are tuned to measurements in \ee{} collisions. The CR tunes introduce new colour reconnection topologies, including junctions, that enhance the baryon production and, to a lesser extent, 
charmonia. 
The three considered tunes (Mode 0, 2, and 3) apply different constraints on the allowed string reconnections, taking into account causal connections of dipoles involved in a reconnection, and time dilation effects caused by relative boosts between string pieces. While multiparton interactions (MPI) are observed in PYTHIA 8 to significantly increase the charm quark production, a modification of the relative abundances of the charm hadron species, with the relative baryon enhancement, is observed only when the MPI are coupled to a color reconnection mode beyond the leading color approximation~\cite{Skands:2014pea}.
It is observed that for the open charm meson ratios the PYTHIA~8 generator predictions with the different tunes are fairly similar and describe the measurements within uncertainty, except for the $\Dplus/\Dzero$ ratio, which is overestimated by about 15\%. However, this difference has a significance of only 1 standard deviation of the combined statistical and systematic uncertainties. 
Significant differences in the PYTHIA~8 predictions are observed when comparing them with the measured baryon-to-meson ratios. The Monash 2013 tune is observed to underestimate the $\Lc/\Dzero$ and $\XicZero/\Dzero$ ratios by nearly $8\,\sigma$ and $2.3\,\sigma$, respectively. It is significantly different from all the CR tunes, which provide an increase of the  baryon-to-meson ratio. Mode 2 is the PYTHIA~8 tune describing the $\Lc/\Dzero$ ratio; however, it still underestimates the $\XicZero/\Dzero$ ratio by about $2\, \sigma$.
For the $\Jpsi/\Dzero$ ratio the CR tunes provide a better description than the Monash 2013 tune. However, all PYTHIA~8 tunes underestimate the measurement. In the simulations, as in the experimental measurement, the \Jpsi{} cross section consists of the prompt and beauty feed-down contributions. 
The fraction of \Jpsi{} from the decay of b-hadrons is about 15\% for $\pt^{\Jpsi{}}>1.3~{\rm GeV}/c$ ~\cite{Chatrchyan:2011kc,Aad:2011sp,Abelev:2012gx}.

In the right panel of Fig.~\ref{figshm}, the measurements are compared with two versions of a statistical hadronisation model (SHM)~\cite{He:2019tik}. One is based on the charm baryon states included by the Particle Data Group (PDG)~\cite{Zyla:2020zbs}, while the other version includes an augmented set of charm baryon states, given by predictions of the relativistic quark model (RQM)~\cite{Ebert:2011kk}. Both versions are reported for two different hadronisation temperatures (\Th)~\cite{He:2019tik}. The two \Th values of 160 MeV and 170 MeV used in the model are above the temperature of 156.5 MeV reported from a fit to the light-flavour hadron yields in central Pb--Pb collisions~\cite{Andronic:2017pug,Cleymans:2020fsc}.
The implementation of the two hadronisation temperatures leads only to small variations in the meson-to-meson ratios, while more significant changes are observed in the baryon sector.
The charm mesons \Dzero, \Dplus, and \Ds{} and baryons are dominantly populated by strong decays from higher-lying charm resonances. Therefore, changes due to an increased temperature on yield ratios relative to \Dzero are due to subtle effects. In particular, in the meson-to-meson ratios a weak sensitivity to temperature and no change due to the added baryons is visible. For the charm baryons, even with the standard PDG spectrum, there is a stronger sensitivity to a temperature increase (dashed and dash-dotted red lines in the right panel of Fig.~\ref{figshm}). The additional baryon states almost double the fraction of the ground-state \Lc{} in the system relative to the PDG scenario, when a hadronisation temperature of 170 MeV is used, and the resulting 
$\Lc/\Dzero$ ratio
becomes comparable to the ALICE measurement~\cite{Acharya:2020uqi}. A similar conclusion is drawn for the production cross section of $\Sigma_\mathrm{c}^{0,+,++}$ baryons in pp collisions at $\sqrt{s}=13~\TeV$~\cite{sigmac}. The $\XicZero/\Dzero$ ratio is observed to increase by a factor 1.3 with respect to the PDG case. With this increase of the \XicZero{} yield, the model calculation is compatible with the measurement within $1.8\, \sigma$. 
No model calculation is available for the $\Jpsi/\Dzero$ ratio.

The \ccbar{} production cross section per unit of rapidity at midrapidity ($\de\sigma^{\ccbar}/\de y|_{|y| < 0.5}$) was calculated by summing the \pt-integrated cross sections of all measured ground-state charm hadrons (\Dzero{}, \Dplus{}, \Ds{}, \Lc{}, and \XicZero{}). The contribution of the \XicZero{} was multiplied by a factor of two, in order to account for the contribution of the \XicPlus{}. The production cross sections of the \XicZero{} and \XicPlus{} baryons were found to be compatible within experimental uncertainties in pp collisions at $\sqrt{s}$ = 13 TeV~\cite{xic13tev}. 
The contribution of \Jpsi{} to the charm production cross section at midrapidity was considered negligible with respect to the other hadron species. Given the absence of measurements of \Omegac{} baryon production at hadron colliders, an asymmetric systematic uncertainty was assigned assuming a contribution equal to the one of \XicZero{} considering the prediction of the Catania model~\cite{Minissale:2020bif}. This uncertainty was summed in quadrature with the other extrapolation uncertainties.
Two correction factors for the different shapes of the rapidity distributions (RS) of charm hadrons 
and ${\ccbar}$ pairs 
were considered. The first factor accounts for the different rapidity distributions of charm hadrons and single charm quarks, and it was evaluated to be unity in the relevant rapidity range based on FONLL calculations. A 2\% uncertainty on this factor was evaluated from the difference obtained with PYTHIA~8. The second correction factor was computed as the ratio $(\de\sigma^\ccbar/\de y) / (\de\sigma^\mathrm{c}/\de y)$, which was estimated from NLO pQCD calculations (POWHEG~\cite{Frixione:2007nw}) to be 1.03. A 3\% uncertainty on this factor was estimated from the difference among the values obtained by varying the factorisation and renormalisation scales independently by a factor of 2 in the POWHEG calculation and using different sets of PDFs (CT10NLO~\cite{Lai:2010vv}, CT14NLO~\cite{Dulat:2015mca}, CT18NNLO~\cite{Hou:2019efy}, CTEQ66~\cite{Nadolsky:2008zw}, and NNPDF31NNLO~\cite{Forte:2020yip}).
The resulting \ccbar{} cross section per unit of rapidity at midrapidity is

\begin{equation}
\frac{\de\sigma^{\ccbar}}{\de y}\bigg{|}_{|y| < 0.5}^{\text{pp, 5.02 TeV}} = 1165\pm44\text{(stat)}^{+ 63}_{- 67}\text{(syst)}^{+ 98}_{- 38}\text{(extr)}\pm43\text{(BR)}\pm42\text{(RS)}\pm24\text{(lumi)}~\mub.
\label{5tev}
\end{equation}

The reported uncertainties in Eq.~\ref{5tev} named (extr) and (BR) refer to extrapolation uncertainties of the charm-hadron cross sections not measured down to \pt = 0 and to the uncertainties of the branching ratios. The extrapolated fraction of the cross section is smaller than 20\%. More details on the extrapolation uncertainties are reported in~\cite{Acharya:2021cqv,Acharya:2020lrg,Acharya:2021dsq}.

The charm fragmentation fractions, $f({\rm c} \rightarrow \Hc)$,  which represent the probabilities of a c quark to hadronise into a given charm hadron, are listed in Table~\ref{tab:FF}. They were obtained by dividing the \pt-integrated cross section of each measured hadron species by the sum of the cross sections of the different ground-state charm hadron species, considering twice the contribution of the \XicZero{} baryon. 
An asymmetric uncertainty to account for the possible sizeable contribution of \Omegac{} was added as done for the evaluation of $\de\sigma^{\ccbar}/\de y$.

\begin{table}[]
\caption{Charm-quark fragmentation fractions into charm hadrons, $f({\rm c} \rightarrow \Hc)$ determined from measurements in pp collisions at $\s = 5.02~\TeV$. Statistical and systematic uncertainties are reported separately. To obtain the complete fragmentation of a c quark, an additional contribution equal to the one of the \XicZero{} should be added to account for the \XicPlus{}. The $f({\rm c} \rightarrow \Lc{})$ includes the feed-down from $\Sigma_{\rm c}^{0,+,++}$ baryons. The sum of the fragmentation fractions adds up to unity within uncertainties, not counting here the $\Dstar$, which feeds into the \Dzero{} and \Dplus{} mesons. }
	\begin{center}
	\renewcommand\arraystretch{1.5}
	\begin{tabular}{cc}
	  $\Hc$  & $f({\rm c} \rightarrow \Hc$)[\%] \\
	  	\hline
	  \Dzero{} &  $39.1 \pm 1.7 ({\rm stat})  _{-3.7}^{+2.5} ({\rm syst}) $  \\
	  \Dplus{} &  $17.3 \pm 1.8 ({\rm stat}) _{-2.1}^{+1.7} ({\rm syst})  $ \\
	  \Ds{} &  $7.3 \pm 1.0 ({\rm stat}) _{-1.1}^{+1.9} ({\rm syst}) $  \\
	  \Lc{} &  $20.4 \pm 1.3 ({\rm stat}) _{-2.2}^{+1.6} ({\rm syst}) $  \\
	  \XicZero{} & $8.0 \pm 1.2 ({\rm stat}) _{-2.4}^{+2.5} ({\rm syst}) $  \\
	\Dstar{} &  $15.5 \pm 1.2 ({\rm stat}) _{-1.9}^{+4.1} ({\rm syst})  $ \\
	\end{tabular}
	\end{center}
	\label{tab:FF}
\end{table}

\begin{figure}[!tb]
\begin{minipage}[c]{0.5\linewidth}
\centering
\includegraphics[width=1.\textwidth]{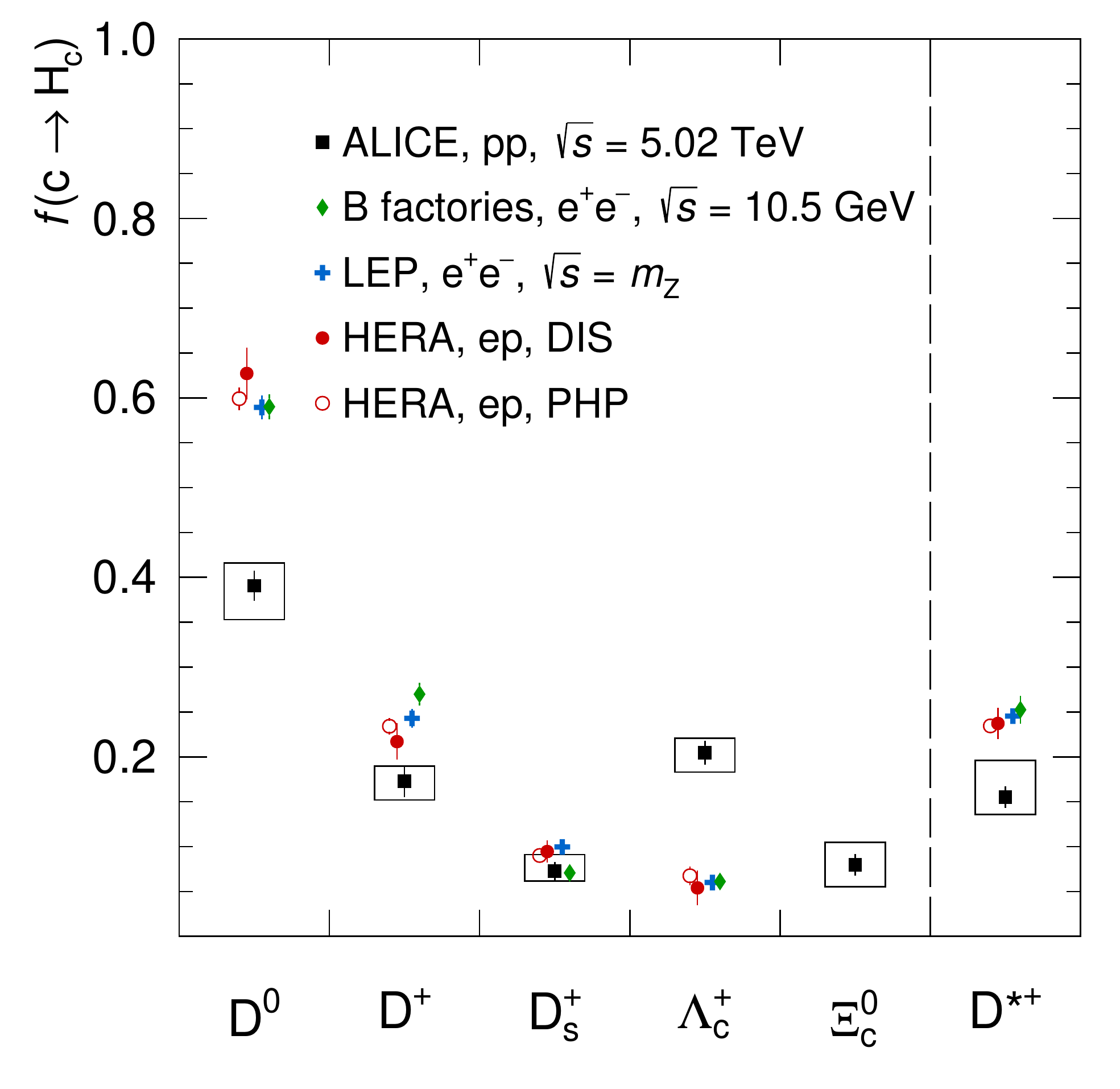}
\end{minipage}
\hspace{0.2cm}
\begin{minipage}[c]{0.5\linewidth}
\centering
\includegraphics[width=1.\textwidth]{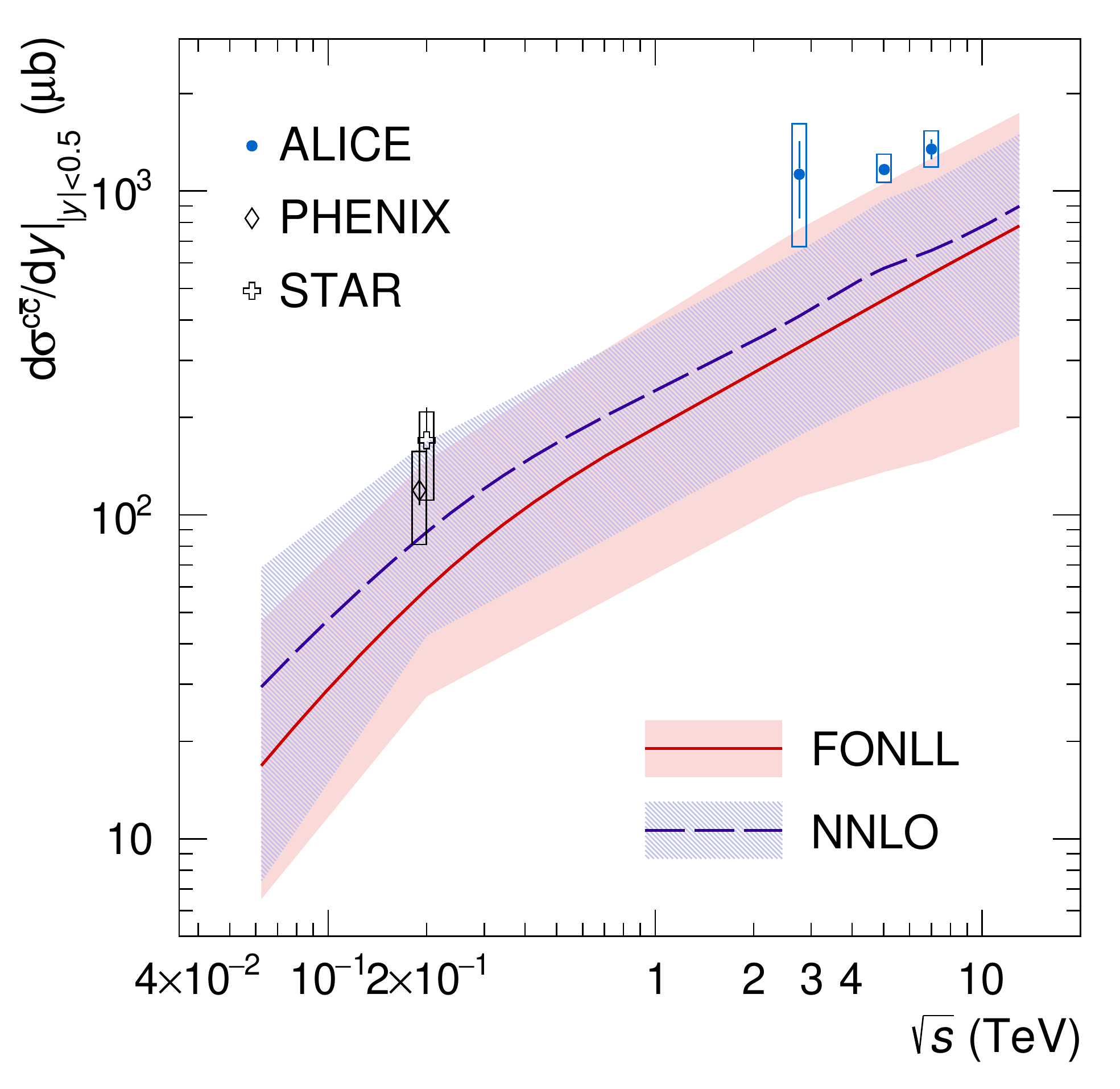}
\end{minipage}%
\caption{Left: Charm-quark fragmentation fractions into charm hadrons measured in pp collisions at $\s = 5.02~\TeV$ in comparison with experimental measurements performed in \ee{} collisions at LEP and at B factories, and in ep collisions at HERA~\cite{Lisovyi:2015uqa}. The \Dstar{} meson is depicted separately since its contribution is also included in the ground-state charm mesons. Right: Charm production cross section at midrapidity per unit of rapidity as a function of the collision energy. STAR~\cite{Adamczyk:2012af} and PHENIX~\cite{Adare:2010de} results, slightly displaced in the horizontal direction for better visibility, are reported. Comparisons with FONLL~\cite{Cacciari:1998it,Cacciari:2001td,Cacciari:2012ny} (red band) and NNLO~\cite{dEnterria:2016ids,dEnterria:2016yhy,Czakon:2013goa} (violet band) pQCD calculations are also shown.}
\label{figxsec}
\end{figure}

In the left panel of Fig.~\ref{figxsec} the fractions $f({\rm c} \rightarrow \Hc)$ are compared with values derived from experimental measurements performed in \ee{} collisions at LEP and B factories as well as in ep collisions~\cite{Lisovyi:2015uqa}. The fragmentation fractions measured at midrapidity in pp collisions at the LHC are different from the ones measured in \ee{} and ep collisions, confirming significant evidence that the assumption of universality (collision-system independence) of parton-to-hadron fragmentation is not valid as reported in~\cite{Acharya:2019mgn,Acharya:2021dsq,Acharya:2020uqi}. The fractions $f({\rm c} \rightarrow \Hc)$ measured in \ee{}, including the \Lc{} baryon, are in agreement with a standard canonical SHM~\cite{Andronic:2009sv}. The $\Lc{}/\Dzero{}$ ratio measured at midrapidity in pp and p--Pb collisions at the LHC is different from the one measured at forward rapidity by the LHCb Collaboration~\cite{Aaij:2013mga,Aaij:2018iyy} as discussed in~\cite{Acharya:2017kfy,Acharya:2020lrg}. 

An increase of about a factor 3.3 for the fragmentation fractions for the \Lc{} baryons with respect to \ee{} and ep collisions, and a concomitant decrease of about a factor 1.4--1.2 for the D mesons, are observed. The significance of the difference considering the uncertainties of both measurements, is about $5\,\sigma$ for \Lc{} baryons. This in turn decreases the fragmentation into \Dzero{} mesons at midrapidity by $6\,\sigma$ with respect to the measurements in \ee{} and ep collisions. In previous measurements in \ee{} and ep collisions no value for the $\XicZero{}$ was obtained and the yield was estimated according to the assumption $f({\rm c} \rightarrow \XicPlus{})/f({\rm c} \rightarrow \Lc{})$ = $f({\rm s} \rightarrow \Xi^{-})/f({\rm s} \rightarrow \Lambda^{0}) \sim$  0.004~\cite{Lisovyi:2015uqa}. The fraction $f({\rm c} \rightarrow \XicZero{})$ was measured for the first time 
and $f({\rm c} \rightarrow \XicZero{})/f({\rm c} \rightarrow \Lc{})$ = $0.39 \pm 0.07(\mathrm{stat})^{+0.08}_{-0.07}(\mathrm{syst})$ was found~\cite{Acharya:2021dsq}. 
A first attempt to compute the fragmentation fractions in pp collisions at the LHC was performed in~\cite{Lisovyi:2015uqa} assuming universal fragmentation, since at that time the measurements of charm baryons at midrapidity were not yet available. The measurements reported here challenge that assumption.

The updated fragmentation fractions obtained for the first time taking into account the measurements of \Dzero{}, \Dplus{}, \Ds{}, \Lc{}, and \XicZero{} at midrapidity in pp collisions at $\s{} = 5.02$ TeV, allowed the recomputation of the charm production cross sections per unit of rapidity at midrapidity in pp collisions at $\s{} = 2.76$ and 7 TeV. The $\Lc/\Dzero$ ratios measured in pp at different collision energies, as well as the $\XicZero/\Dzero$ ratio, are compatible~\cite{Acharya:2021dsq,sigmac,Acharya:2020lrg}.
The charm cross sections were obtained by scaling the \pt-integrated \Dzero{}-meson cross section~\cite{Abelev:2012vra,Acharya:2017jgo} for the relative fragmentation fraction of a charm quark into a \Dzero{} meson measured in pp collisions at $\s = 5.02~\TeV$ and applying the two correction factors for the different shapes of the rapidity distributions of charm hadrons and $\rm c\bar{c}$ pairs. The \pt-integrated \Dzero{}-meson cross section was used because at the other energies not all charm hadrons were measured and the \Dzero{} measurements are the most precise. The uncertainties of the fragmentation fraction (FF) were taken into account in calculating the \ccbar{} production cross section as was the uncertainty introduced by the rapidity correction factors. The BR of the \DzerotoKpi{} decay channel was also updated, considering the latest value reported in the PDG~\cite{Zyla:2020zbs}.

The resulting \ccbar{} cross sections per unit of rapidity at midrapidity are

\begin{equation}
\frac{\de\sigma^{\ccbar}}{\de y}\bigg{|}_{|y| < 0.5}^{\text{pp, 7 TeV}} = 1347\pm97\text{(stat)}\pm104\text{(syst)} \pm 11\text{(BR)} \pm _{-105}^{+142}\text{(FF)}\pm44\text{(RS)}\pm47\text{(lumi)}~\mub,
	\label{totalxsec}
\end{equation}

and
\begin{equation}
\frac{\de\sigma^{\ccbar}}{\de y}\bigg{|}_{|y| < 0.5}^{\text{pp, 2.76 TeV}} = 1126\pm303\text{(stat)}^{+258}_{-429}\text{(syst)}^{+397}_{-53}\text{(extr)} \pm 9\text{(BR)} \pm ^{+119}_{-88}\text{(FF)}\pm61\text{(RS)}\pm21\text{(lumi)}~\mub
	\label{totalxsec276}
\end{equation}
for $\s = 7$ and 2.76 TeV, respectively. The updated \ccbar{} cross sections at $\s{} = 2.76$ and 7 TeV are about 40\% higher than the previously published results~\cite{Abelev:2012vra,Acharya:2017jgo}, reflecting the differences in the fragmentation into charm baryons measured in \ee{} and pp collisions.

In the right panel of Fig.~\ref{figxsec}, the measured \ccbar{} cross sections are compared with FONLL and NNLO predictions as a function of the collision energy. The NNLO values were obtained by the authors of~\cite{dEnterria:2016ids,dEnterria:2016yhy} by applying to the central value of the FONLL ${\rm d}\sigma^{\ccbar}/{\rm d}y$ a $K$ factor (NNLO/NLO) calculated with a modified version of the top++ code~\cite{Czakon:2013goa} with parameter values as in~\cite{dEnterria:2016ids,dEnterria:2016yhy} and using the relative scale uncertainties obtained at NNLO with top++.
The \ccbar{} cross sections are also compared with the STAR~\cite{Adamczyk:2012af} and PHENIX~\cite{Adare:2010de} results measured in pp collisions at $\s = 200~\GeV$. 
The STAR measurement is obtained by scaling the \Dzero{} and \Dstar{} cross sections by the charm-quark fragmentation fractions measured in \ee{} collisions from the CLEO and BELLE experiments~\cite{Lisovyi:2015uqa}.
The PHENIX \ccbar{} cross section is obtained from the measurement of the cross sections of electrons from semileptonic heavy-flavour hadron decays. Both results are compatible within uncertainties with the upper edge of the FONLL and NNLO bands.
The \ccbar{} cross sections measured at the three LHC collision energies are higher than the upper edge of the FONLL and NNLO bands; however, they are compatible within approximately one standard deviation of the experimental uncertainty. The theoretical uncertainties are estimated as a convolution of the pQCD calculations obtained by varying the factorisation and renormalisation scales. The uncertainties of the PDFs and of the charm-quark mass are also included in the uncertainties of both calculations and are determined with FONLL as described in~\cite{Cacciari:2012ny}.

In summary, the charm production cross section per unit of rapidity at midrapidity in pp collisions at $\s = 5.02~\TeV$ was determined by exploiting recent measurements of the ground-state charm hadrons, including for the first time the measured baryon states. 
The charm fragmentation fractions $f({\rm c} \rightarrow \Hc)$ were computed for the first time in hadron collisions at the LHC using measurements of charm baryons at midrapidity, and they were found to be different from those measured in \ee{} and ep collisions. This observation indicates that the hadronisation of charm quarks into charm hadrons is not a universal process among different collision systems. The fragmentation fraction for the \XicZero{} baryon was measured for the first time and found to be sizeable.
Finally, the charm production cross section per unit of rapidity at midrapidity, in pp collisions at $\sqrt{s} = 5.02$ TeV at the LHC was measured and lies at the upper edge of the theoretical pQCD calculations.


\newenvironment{acknowledgement}{\relax}{\relax}
\begin{acknowledgement}
\section*{Acknowledgements}
\input{fa_2021-05-05.tex}
\end{acknowledgement}

\bibliographystyle{utphys}   
\bibliography{bibliography}

\newpage
\appendix

%
%

\section{The ALICE Collaboration}
\label{app:collab}
\input{2021-05-05-Alice_Authorlist_2021-05-05.tex}  
\end{document}

%% file: commands.tex
%

\newcommand{\pp}           {pp\xspace}
\newcommand{\ppbar}        {\mbox{$\mathrm {p\overline{p}}$}\xspace}
\newcommand{\XeXe}         {\mbox{Xe--Xe}\xspace}
\newcommand{\PbPb}         {\mbox{Pb--Pb}\xspace}
\newcommand{\pA}           {\mbox{pA}\xspace}
\newcommand{\pPb}          {\mbox{p--Pb}\xspace}
\newcommand{\AuAu}         {\mbox{Au--Au}\xspace}
\newcommand{\dAu}          {\mbox{d--Au}\xspace}
\newcommand{\ee}           {\ensuremath{\mathrm{e^+e^-}}\xspace}

\newcommand{\s}            {\ensuremath{\sqrt{s}}\xspace}
\newcommand{\snn}          {\ensuremath{\sqrt{s_{\mathrm{NN}}}}\xspace}
\newcommand{\pt}           {\ensuremath{p_{\rm T}}\xspace}
\newcommand{\meanpt}       {$\langle p_{\mathrm{T}}\rangle$\xspace}
\newcommand{\ycms}         {\ensuremath{y_{\rm CMS}}\xspace}
\newcommand{\ylab}         {\ensuremath{y_{\rm lab}}\xspace}
\newcommand{\etarange}[1]  {\mbox{$\left | \eta \right |~<~#1$}}
\newcommand{\yrange}[1]    {\mbox{$\left | y \right |~<~#1$}}
\newcommand{\dndy}         {\ensuremath{\mathrm{d}N_\mathrm{ch}/\mathrm{d}y}\xspace}
\newcommand{\dndeta}       {\ensuremath{\mathrm{d}N_\mathrm{ch}/\mathrm{d}\eta}\xspace}
\newcommand{\avdndeta}     {\ensuremath{\langle\dndeta\rangle}\xspace}
\newcommand{\dNdy}         {\ensuremath{\mathrm{d}N_\mathrm{ch}/\mathrm{d}y}\xspace}
\newcommand{\Npart}        {\ensuremath{N_\mathrm{part}}\xspace}
\newcommand{\Ncoll}        {\ensuremath{N_\mathrm{coll}}\xspace}
\newcommand{\dEdx}         {\ensuremath{\textrm{d}E/\textrm{d}x}\xspace}
\newcommand{\RpPb}         {\ensuremath{R_{\rm pPb}}\xspace}
\newcommand{\Tfo}         {\ensuremath{T_\mathrm{fo}}\xspace}
\newcommand{\Th}         {\ensuremath{T_\mathrm{h}}\xspace}
\newcommand{\de}          {\ensuremath{\mathrm{d}}\xspace}

\newcommand{\nineH}        {$\sqrt{s}~=~0.9$~Te\kern-.1emV\xspace}
\newcommand{\seven}        {$\sqrt{s}~=~7$~Te\kern-.1emV\xspace}
\newcommand{\twoH}         {$\sqrt{s}~=~0.2$~Te\kern-.1emV\xspace}
\newcommand{\twosevensix}  {$\sqrt{s}~=~2.76$~Te\kern-.1emV\xspace}
\newcommand{\five}         {$\sqrt{s}~=~5.02$~Te\kern-.1emV\xspace}
\newcommand{\twosevensixnn}{$\sqrt{s_{\mathrm{NN}}}~=~2.76$~Te\kern-.1emV\xspace}
\newcommand{\fivenn}       {$\sqrt{s_{\mathrm{NN}}}~=~5.02$~Te\kern-.1emV\xspace}
\newcommand{\LT}           {L{\'e}vy-Tsallis\xspace}
\newcommand{\GeVc}         {\ensuremath{\mathrm{GeV}/c}\xspace}
\newcommand{\MeVc}         {\ensuremath{\mathrm{MeV}/c}\xspace}
\newcommand{\TeV}          {\ensuremath{\mathrm{TeV}}\xspace}
\newcommand{\GeV}          {\ensuremath{\mathrm{GeV}}\xspace}
\newcommand{\MeV}          {\ensuremath{\mathrm{MeV}}\xspace}
\newcommand{\GeVmass}      {\ensuremath{\mathrm{GeV}/c^2}\xspace}
\newcommand{\MeVmass}      {\ensuremath{\mathrm{MeV}/c^2}\xspace}
\newcommand{\lumi}         {\ensuremath{\mathcal{L}}\xspace}
\newcommand{\mub}         {\ensuremath{\mu\mathrm{b}}\xspace}

\newcommand{\ITS}          {\rm{ITS}\xspace}
\newcommand{\TOF}          {\rm{TOF}\xspace}
\newcommand{\ZDC}          {\rm{ZDC}\xspace}
\newcommand{\ZDCs}         {\rm{ZDCs}\xspace}
\newcommand{\ZNA}          {\rm{ZNA}\xspace}
\newcommand{\ZNC}          {\rm{ZNC}\xspace}
\newcommand{\SPD}          {\rm{SPD}\xspace}
\newcommand{\SDD}          {\rm{SDD}\xspace}
\newcommand{\SSD}          {\rm{SSD}\xspace}
\newcommand{\TPC}          {\rm{TPC}\xspace}
\newcommand{\TRD}          {\rm{TRD}\xspace}
\newcommand{\VZERO}        {\rm{V0}\xspace}
\newcommand{\VZEROA}       {\rm{V0A}\xspace}
\newcommand{\VZEROC}       {\rm{V0C}\xspace}
\newcommand{\Vdecay} 	   {\ensuremath{V^{0}}\xspace}

\newcommand{\pip}          {\ensuremath{\pi^{+}}\xspace}
\newcommand{\pim}          {\ensuremath{\pi^{-}}\xspace}
\newcommand{\kap}          {\ensuremath{\rm{K}^{+}}\xspace}
\newcommand{\kam}          {\ensuremath{\rm{K}^{-}}\xspace}
\newcommand{\pbar}         {\ensuremath{\rm\overline{p}}\xspace}
\newcommand{\kzero}        {\ensuremath{{\rm K}^{0}_{\rm{S}}}\xspace}
\newcommand{\lmb}          {\ensuremath{\Lambda}\xspace}
\newcommand{\almb}         {\ensuremath{\overline{\Lambda}}\xspace}
\newcommand{\Om}           {\ensuremath{\Omega^-}\xspace}
\newcommand{\Mo}           {\ensuremath{\overline{\Omega}^+}\xspace}
\newcommand{\X}            {\ensuremath{\Xi^-}\xspace}
\newcommand{\Ix}           {\ensuremath{\overline{\Xi}^+}\xspace}
\newcommand{\Xis}          {\ensuremath{\Xi^{\pm}}\xspace}
\newcommand{\Oms}          {\ensuremath{\Omega^{\pm}}\xspace}
\newcommand{\degree}       {\ensuremath{^{\rm o}}\xspace}
\newcommand{\Hc}           {\ensuremath{\mathrm{H_c}}\xspace}
\newcommand{\Dzero}        {\ensuremath{\mathrm{D^0}}\xspace}
\newcommand{\Dplus}        {\ensuremath{\mathrm{D^+}}\xspace}
\newcommand{\Dstar}        {\ensuremath{\mathrm{D^{*+}}}\xspace}
\newcommand{\Ds}           {\ensuremath{\mathrm{D_s^+}}\xspace}
\newcommand{\Lc}           {\ensuremath{\Lambda_\mathrm{c}^+}\xspace}
\newcommand{\XicZero}      {\ensuremath{\Xi_\mathrm{c}^0}\xspace}
\newcommand{\XicPlus}      {\ensuremath{\Xi_\mathrm{c}^+}\xspace}
\newcommand{\XicPlusZero}  {\ensuremath{\Xi_\mathrm{c}^{+,0}}\xspace}
\newcommand{\Omegac}       {\ensuremath{\Omega_\mathrm{c}^0}\xspace}
\newcommand{\Jpsi}         {\ensuremath{\mathrm{J}/\psi}\xspace}
\newcommand{\ccbar}        {\ensuremath{\mathrm{c\overline{c}}}\xspace}
\newcommand{\bbbar}        {\ensuremath{\mathrm{b\overline{b}}}\xspace}
\newcommand{\Lb}           {\ensuremath{\Lambda_\mathrm{b}^+}\xspace}
\newcommand{\Xib}          {\ensuremath{\Xi_\mathrm{b}^{0,-}}\xspace}
\newcommand{\DzerotoKpi}   {\ensuremath{\mathrm{D^0\to K^-\pi^+}}}

%% file: fa_2021-05-05.tex

The ALICE Collaboration would like to thank all its engineers and technicians for their invaluable contributions to the construction of the experiment and the CERN accelerator teams for the outstanding performance of the LHC complex.
The ALICE Collaboration gratefully acknowledges the resources and support provided by all Grid centres and the Worldwide LHC Computing Grid (WLCG) collaboration.
The ALICE Collaboration acknowledges the following funding agencies for their support in building and running the ALICE detector:
A. I. Alikhanyan National Science Laboratory (Yerevan Physics Institute) Foundation (ANSL), State Committee of Science and World Federation of Scientists (WFS), Armenia;
Austrian Academy of Sciences, Austrian Science Fund (FWF): [M 2467-N36] and Nationalstiftung f\"{u}r Forschung, Technologie und Entwicklung, Austria;
Ministry of Communications and High Technologies, National Nuclear Research Center, Azerbaijan;
Conselho Nacional de Desenvolvimento Cient\'{\i}fico e Tecnol\'{o}gico (CNPq), Financiadora de Estudos e Projetos (Finep), Funda\c{c}\~{a}o de Amparo \`{a} Pesquisa do Estado de S\~{a}o Paulo (FAPESP) and Universidade Federal do Rio Grande do Sul (UFRGS), Brazil;
Ministry of Education of China (MOEC) , Ministry of Science \& Technology of China (MSTC) and National Natural Science Foundation of China (NSFC), China;
Ministry of Science and Education and Croatian Science Foundation, Croatia;
Centro de Aplicaciones Tecnol\'{o}gicas y Desarrollo Nuclear (CEADEN), Cubaenerg\'{\i}a, Cuba;
Ministry of Education, Youth and Sports of the Czech Republic, Czech Republic;
The Danish Council for Independent Research | Natural Sciences, the VILLUM FONDEN and Danish National Research Foundation (DNRF), Denmark;
Helsinki Institute of Physics (HIP), Finland;
Commissariat \`{a} l'Energie Atomique (CEA) and Institut National de Physique Nucl\'{e}aire et de Physique des Particules (IN2P3) and Centre National de la Recherche Scientifique (CNRS), France;
Bundesministerium f\"{u}r Bildung und Forschung (BMBF) and GSI Helmholtzzentrum f\"{u}r Schwerionenforschung GmbH, Germany;
General Secretariat for Research and Technology, Ministry of Education, Research and Religions, Greece;
National Research, Development and Innovation Office, Hungary;
Department of Atomic Energy Government of India (DAE), Department of Science and Technology, Government of India (DST), University Grants Commission, Government of India (UGC) and Council of Scientific and Industrial Research (CSIR), India;
Indonesian Institute of Science, Indonesia;
Istituto Nazionale di Fisica Nucleare (INFN), Italy;
Institute for Innovative Science and Technology , Nagasaki Institute of Applied Science (IIST), Japanese Ministry of Education, Culture, Sports, Science and Technology (MEXT) and Japan Society for the Promotion of Science (JSPS) KAKENHI, Japan;
Consejo Nacional de Ciencia (CONACYT) y Tecnolog\'{i}a, through Fondo de Cooperaci\'{o}n Internacional en Ciencia y Tecnolog\'{i}a (FONCICYT) and Direcci\'{o}n General de Asuntos del Personal Academico (DGAPA), Mexico;
Nederlandse Organisatie voor Wetenschappelijk Onderzoek (NWO), Netherlands;
The Research Council of Norway, Norway;
Commission on Science and Technology for Sustainable Development in the South (COMSATS), Pakistan;
Pontificia Universidad Cat\'{o}lica del Per\'{u}, Peru;
Ministry of Education and Science, National Science Centre and WUT ID-UB, Poland;
Korea Institute of Science and Technology Information and National Research Foundation of Korea (NRF), Republic of Korea;
Ministry of Education and Scientific Research, Institute of Atomic Physics and Ministry of Research and Innovation and Institute of Atomic Physics, Romania;
Joint Institute for Nuclear Research (JINR), Ministry of Education and Science of the Russian Federation, National Research Centre Kurchatov Institute, Russian Science Foundation and Russian Foundation for Basic Research, Russia;
Ministry of Education, Science, Research and Sport of the Slovak Republic, Slovakia;
National Research Foundation of South Africa, South Africa;
Swedish Research Council (VR) and Knut \& Alice Wallenberg Foundation (KAW), Sweden;
European Organization for Nuclear Research, Switzerland;
Suranaree University of Technology (SUT), National Science and Technology Development Agency (NSDTA) and Office of the Higher Education Commission under NRU project of Thailand, Thailand;
Turkish Energy, Nuclear and Mineral Research Agency (TENMAK), Turkey;
National Academy of  Sciences of Ukraine, Ukraine;
Science and Technology Facilities Council (STFC), United Kingdom;
National Science Foundation of the United States of America (NSF) and United States Department of Energy, Office of Nuclear Physics (DOE NP), United States of America.

%% file: 2021-05-05-Alice_Authorlist_2021-05-05.tex
\begin{flushleft} 

S.~Acharya$^{\rm 143}$, 
D.~Adamov\'{a}$^{\rm 98}$, 
A.~Adler$^{\rm 76}$, 
J.~Adolfsson$^{\rm 83}$, 
G.~Aglieri Rinella$^{\rm 35}$, 
M.~Agnello$^{\rm 31}$, 
N.~Agrawal$^{\rm 55}$, 
Z.~Ahammed$^{\rm 143}$, 
S.~Ahmad$^{\rm 16}$, 
S.U.~Ahn$^{\rm 78}$
% ...

\bigskip

\bigskip 

\textbf{\Large Affiliation Notes}

\bigskip 

% ...
$^{\rm V}$ Also at: Institute of Theoretical Physics, University of Wroclaw, Poland\\

\bigskip

\bigskip 

\textbf{\Large Collaboration Institutes}

\bigskip 

% ...
$^{16}$ Department of Physics, Aligarh Muslim University, Aligarh, India\\
% ...
$^{31}$ Dipartimento DISAT del Politecnico and Sezione INFN, Turin, Italy\\
% ...
$^{35}$ European Organization for Nuclear Research (CERN), Geneva, Switzerland\\
% ...
$^{55}$ INFN, Sezione di Bologna, Bologna, Italy\\
% ...
$^{76}$ Johann-Wolfgang-Goethe Universit\"{a}t Frankfurt Institut f\"{u}r Informatik, Fachbereich Informatik und Mathematik, Frankfurt, Germany\\
% ...
$^{78}$ Korea Institute of Science and Technology Information, Daejeon, Republic of Korea\\
% ...
$^{83}$ Lund University Department of Physics, Division of Particle Physics, Lund, Sweden\\
% ...
$^{98}$ Nuclear Physics Institute of the Czech Academy of Sciences, \v{R}e\v{z} u Prahy, Czech Republic\\
% ...
$^{143}$ Variable Energy Cyclotron Centre, Homi Bhabha National Institute, Kolkata, India\\
% ...

\end{flushleft} 